\documentclass[reprint,amsmath,amssymb,aps,prl,floats]{revtex4-1}
\usepackage{graphicx}
\usepackage{dcolumn}
\usepackage{bm}

\begin{document}
\preprint{RSI/PMT}

\title{MW-Optical Double Resonance in $^{171}{Yb}^+$ Trapped Single Ion and its Application for Precision Experiments}

\author{S. Rahaman}\email{saidur.rahaman@physics.gatech.edu}
\author{J. Danielson}
\author{M. Schacht}\email{mchacht@lanl.gov}
\author{M. Schauer}
\author{J. Zhang}
\author{J. Torgerson}\thanks{Present address: Quantel Medical Unit, Leader of Solid State Lasers, Bozeman, Montana-59715, USA}
\affiliation{Physics Division, P-23, Los Alamos National Laboratory, Mail Stop: H803}

\collaboration{Los Alamos Unclassified Release \#:
LA-UR 11-06676.}

\begin{abstract}
We have employed the 12.6 GHz microwave transition resonance of a single trapped$^{171}$Yb$+$ ion to accurately measure the size and relative orientation of the magnetic and optical electric fields at the position of the ion in the trap. Accurate knowledge of these fields is required for precision experiments such as single ion PNC. As a proof of the principle we have measured the polarization dependent light-shift of the ground state hyperfine levels due to the 369nm cooling laser to determine its electric field amplitude and polarization.

\begin{description}
\item[Usage] www.arXiv.org.
\item[PACS numbers] 32.80.Pj,32.80Ps, 42.62.-b,42.62.Fi,33.60, 32.10. Fn
\end{description}

\end{abstract}

\maketitle

\section{\label{sec:Introduction}Introduction}

The 12.6 GHz ground state hyperfine transition in $^{171}$Yb$+$ ion is attractive as a practical microwave frequency standard. Several experiments have measured this transition frequency very accurately {\cite{Warrington2002}} allowing it to be used as a sensitive probe of the E+M environment of the ion. In particular, the magnetic field and the amplitude and polarization of optical fields that couple to the ground state can be
measured from the light shifts of these ground state hyperfine levels.

Accurate knowledge of these fields is especially important in precision measurements such as atomic parity violation {\cite{Fortson1993}} in which information about beyond Standard Model physics appears as spin dependent light shifts of states coupled by strong optical fields. These shifts are proportional to a parity violation induced mixing of atomic states and the size of the applied electric field. Extracting a precise measurement of this mixing then requires equally precise knowledge of the size of the field. This is difficult to determine indirectly from laser power and beam geometry due to large uncertainties in the losses and aberrations of the beam between the ion and where the beam is accessible and can be measured, and the precise position of the ion in the beam. This has been cited as a distinct disadvantage of Ion PNC experiments compared to other Atomic Parity Violation experiments where fields can be easily determined from applied voltages and electrode geometry {\cite{Mandel2010}}. But, measuring the spin independent component of the light shift using the ground state hyperfine shift allows a direct and precise measurement of the applied electric field that, in fact, improves on previous experiments by eliminating uncertainties due to non-uniform fields from fringing effects and electrode imperfections.

The Ion PNC experiment is also subject to possible large systematic errors due to imperfections in the polarization of the the applied electric fields. For a particular ideal case of linearly polarized light, a small spurious circular polarization can give a spin dependent shift that is larger than that due to
parity violation. The additional information from the light shifts of all the ground state hyperfine sub-levels can be used as independent information about the polarization to reduce and eliminate or correct for the resulting PNC mimicking errors.

We have developed these methods and applied them to the case of light shifts due to the 369nm cooling laser and measured via the 12.6GHz hyperfine transition. These yield a measurement of the 369nm electric field amplitude
limited by laser intensity stability and atomic theory, and qualitative determination of the polarization of this field relative to an applied magnetic field, also limited by 369nm laser intensity stability and, in this
case, polarization stability.

\section{$^{171}$Yb$+$}

\begin{figure}[h]
\resizebox{\columnwidth}{!}{\includegraphics{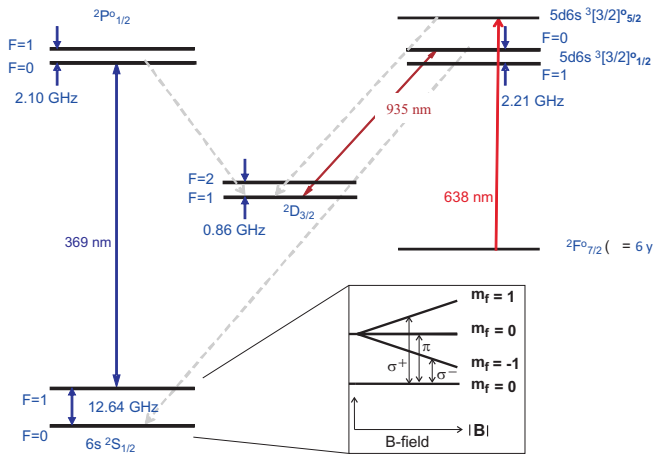}}
\caption{\label{fig:EnergyLevelDiagram}Energy level diagram of $^{171}$Yb$^+$ (not to scale) showing the hyperfine levels, transition wavelengths and lifetime of the states of interest. Inset shows the Zeeman structure of $^2 S_{1/2}$ state in the presence of a very low magnetic field and possible $\pi$ and $\sigma_{\pm}$ microwave transitions between these levels.}
\end{figure}

A partial energy level diagram for $^{171}$Yb$+$ is shown in Fig. \ref{fig:EnergyLevelDiagram}. Doppler cooling and state detection is done using the 369nm transition between the ground $^2 S_{1/2}$(F=1) and excited $^2 P_{1/2}$(F=0) which has a line-width of 23 MHz. The ground and the excited states of $^{171}$Yb$+$ are split into F=0,1 hyperfine multiplets with energy differences of 12.6 GHz and 2.10 GHz. The $^2 S_{1/2}$(F=1) to $^2
P_{1/2}$(F=1) transition will have a small off-resonant coupling that would result in optical pumping from decay into the $^2 S_{1/2}$(F=0) state. To avoid this the 369nm laser is phase modulated at 14.7 GHz with an external tuned electro-optic modulator (EOM). This couples the ground state F=0 level to the P state F=1 level and the ion re-enters the cooling cycle through other transitions.

Population in the $^2 P_{1/2}$ decays primarily back to the $^2 S_{1/2}$ state, but also to the metastable $^2 D_{3/2}$ state ($\tau$ = 50 ms) with a branching ratio of about 1:200. The population in the $^2 D_{3/2}$ is returned to the ground state through the $^2 D_{3/2}$ - $^3 3/2_{1/2}$ transition at 935 nm. A 3.1 GHz EOM is used to generate sidebands that will couple both $D$ hyperfine levels to a state that will decay back to the ground state to resume cooling.

Laser intensities similar to that sufficient to saturate both of these transitions are used. The florescence from $P \rightarrow S$ decays are detected with a PMT and counted. A maximum count rate of about $10^4$ counts/s
indicates the presence of an ion in any state that is part of the cooling cycle.

The inset in Fig. 1 shows the energy levels of $^{171}$Yb$+$ which are relevant for the RF-optical double resonance. The largest contributions to the energies of these level are the linear Zeeman shift due to a static magnetic field, and a polarization dependent light shift from the 369nm cooling laser. These shifts will split the three $F=1$ sub-levels and the final energies can be used to determine the size of the magnetic field, and the amplitude and polarization of the laser electric field.

A conventional linear RF trap was used for this experiment. It consists of four rods about 50 mm long and center to the rod distance of about 0.8 mm. This trap is driven at a frequency of $\Omega=15.6$ MHz with about $V=700$ volts. This gives a pseudo-potential depth of $U \approx 100$ eV, yielding a secular frequency $f_{sec} = 500$ kHz and an axial frequency = 100 kHz for a Yb$^+$ ion. A more detailed description of this trap system can be found in {\cite{Schauer2010}}.

To measure the hyperfine energy levels, the 14 GHz EOM is turned off during cooling. This results in the ion being quickly driven to the $F=0$ hyperfine level of the ground state and stopping the florescence associated with the ion state moving through the cooling signal. The PMT signal then decays to a low background rate of a few $100$cps in $<$ 1ms.


The 12.4 GHz transition is then driven directly using a frequency synthesizer amplified to between 0-10 dBm and output to a microwave horn placed about 10cm from the trap and directed at the trap center through the same window used by the PMT to detect the ion. When the driving frequency coincides with a transition energy to one of the $F=1$ hyperfine levels, the ion is driven back into a state accessible by the cooling beams resulting in an increase in the PMT count rate as the ion is again able to fluoresce after excitations to the $P$ state. The amount of the increase is determined by the hyperfine transition rate. At the highest MW powers used, on resonance the resulting PMT signal is 1/4 to 1/2 of the rate observed when the 14 GHz EOM is active. This suggests that the MW driven hyperfine transition rate is the same order of magnitude as the pumping rate, $\gtrsim 1 \text{kHz}$, as the ion spends only a short time in the $F=0$ state.

If no $F=1$ are degenerate resonances will appear at the three frequencies corresponding to a transition between the $F=0$ state and a particular $F=1$ state. Measuring this rate as a function of MW drive frequency allows determining the frequencies of $F=0\rightarrow F=1$ transitions. A typical result is shown in Fig.~2. The linewidth of the peak is about 0.3 MHz and the resulting sensitivity of the frequency measurement is about 2.3$\times10^{-5}$.

\begin{figure}[h]
\resizebox{\columnwidth}{!}{
\includegraphics{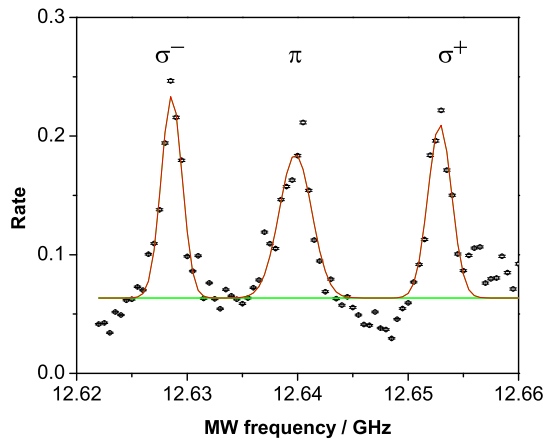}}
\caption{\label{fig:2} Hyperfine transition resonances in the $^2 S_{1/2}$ state showing light shifts due to linearly polarized 369nm light at 45 degree angles.}
\end{figure}


\section{Zeeman Shifts and Magnetic Fields}

In the absence of optical fields the energies of the $F=1$ levels will be determined by the external applied static magnetic field. For a particular $F=1$ sublevel having $j_z=m$ these energies are largely determined by the linear Zeeman shift: \[ \Delta \omega_m = \frac{\Delta E_m}{h} = m \frac{\mu_B}{h} B = m 1.4 \text{MHz} \left( B / \text{Gauss} \right) \]
Quadratic corrections to this from hyperfine couplings will be of order
$$\Delta\omega_m^2/\Omega_{hf}\sim 10 \text{MHz}(10 \text{MHz}/10 \text{GHz})=10 \text{kHz}$$
and are negligible for the precision required here and not detectable with the linewidths involved in this measurement method.

\begin{figure}[h]
\resizebox{\columnwidth}{!}{\includegraphics{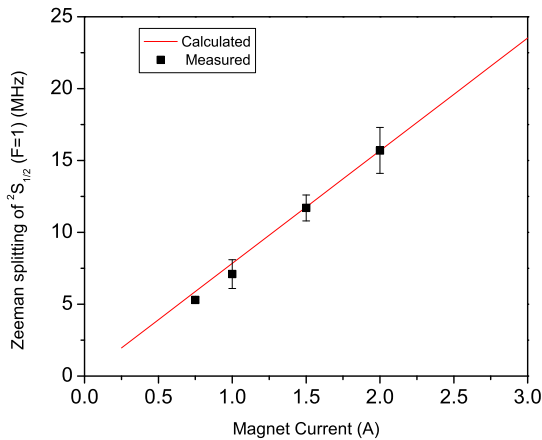}}
\caption{\label{fig:zeemanPlot}Zeeman splitting of the $^2 S_{1/2}$(F=1) level as a function of the current flowing in the Helmholtz coil for generating the low magnetic field. The solid line represents the calculated difference in Zeeman energy splitting between the $m_f=-1$ to $m_f=1$ levels in MHz. The filled squares were the measured difference in Zeeman energy splitting of the same states.}
\end{figure}

Only the $m=\pm 1$ states are shifted. These shifts can then be used to determine the size of the magnetic field at the ion. Figure \ref{fig:zeemanPlot} shows $\Delta \omega_{+1} - \Delta \omega_0$ as a functions of current applied to a Helmholtz coil. The solid line shows the shift expected from the magnetic field calculated using the geometry of the coil to be 5.6 Gauss/Amp. This data was taken with the polarization of the cooling beams parallel to the magnetic field to minimize the light shifts of these states, as further discussed below, so that only the Zeeman shifts affect the resonance frequencies.

The magnetic field strength can be determined as accurately as the resonance frequency of the transition can be measured. In this case $\sim 10 \text{MHz}$ shifts are determined to about $10\%$. This can be improved with increased data collection time and made more accurate than the value determined from coil current and geometry as it naturally include offsets from the Earth's field and nearby magnetic materials. If using this double resonance method, care must be taken to insure that the levels are not also shifted by the optical fields used to probe the population of the $^2 S_{1/2}$(F=1) level. This is difficult to do independently to better than a few percent so for more precise requirements another method much be used that drive and detect the hyperfine transitions when no optical fields are present. However this method has a much simpler measurement sequence and significantly larger sensitivity.


\section{Light Shifts}

The 369nm cooling laser also affects the energies of these levels through the AC Stark effect. The simplest case is that of only two coupled states, as for linearly polarized light parallel to the magnetic field which drives only $\Delta m=0$ transitions. On resonance, using the rotating wave approximation, the shift is given by the size of the dipole coupling between the two states $\delta\omega=\Omega/h$ where
\begin{align*}
\Omega &=e \frac{E}{2}\left<S,m=0\left|z\right|P,m=0\right>
\end{align*}

The Wigner-Eckhart theorem gives the matrix element in terms of the Clebsch-Gordan coefficient $\left<1,0;1,0,0,0\right>=-1/\sqrt{3}$ and the reduced matrix element $\left<S\left|\left|r\right|\right|P\right>$. The reduced matrix element can be determined from the P state lifetime, $2\pi\tau=1/\Gamma$, from
$$\Gamma=\frac{4\alpha}{3}\frac{\omega^3}{c^2}\frac{\left|\left<S\left|\left|r\right|\right|P\right>\right|^2}{2 j_S+1}$$
With $\tau\sim 8ms$, $\left<S\left|\left|r\right|\right|P\right>\sim 0.037 nm$ giving
$$\delta\omega=-2.6 \text{kHz}\thinspace E/(V/m)$$
For laser powers of $P\sim 50\mu W$ and spot sizes of $\sigma\sim 100 \mu m$, $E\sim 2000 V/m$, giving shifts on the order of
$$\delta\omega\sim 5-6 \text{MHz}$$
The size of the shift can then be used to determine the electric field.

\begin{figure}[h]
\resizebox{\columnwidth}{!}{\includegraphics{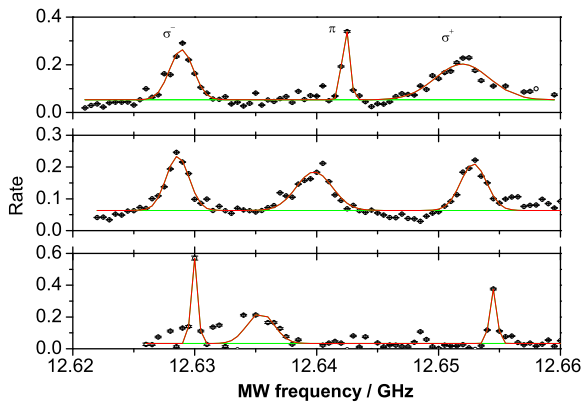}}
\caption{\label{fig:MWTransitions}
Hyperfine transition resonances in the $^2 S_{1/2}$ state showing light shifts due to linearly polarized 369nm light at various angles. Top: $\vec E\perp\vec B$ Center: $\angle\vec E\vec B=45^\circ$ Bottom: $\vec E\parallel\vec B$
}
\end{figure}

Figure \ref{fig:MWTransitions} shows the resulting shift from this configuration. The shift of the $\Delta m=0$ transition resonance is $-6.70\pm 0.12\text{MHz}$, corresponding to an electric field of $E=2580\pm 46 \text{V/cm}$, a relative precision of $1.8\%$.

Note that the $\Delta m=0$ peak is considerable broader then the $\Delta m=\pm1$ peaks. The corresponds to intensity fluctuations on the 369nm laser. The size of the shift is proportional to the electric field so changes in the laser power move the resonance and at long timescales this appears as a broadening of transition line. Long timescale changes in the laser power were reduced to $5-6\%$ manually using a variable attenuator and a power meter. The width of the $\Delta m=0$ line is $\sim 1\text{MHz}$ or $15\%$ of the size of the shift. This is considerable larger than the long time, time averaged variations suggesting shorter term fluctuations on the order of $30\%$ consistent with the limits of the locking of the doubling cavity used to generate this 369nm light. The $\Delta m=\pm 1$ peaks remain sharp confirming that they are largely uncoupled in this geometry, the show no shift and no broadening.

For arbitrary tunings the shift is also frequency dependent having the usual Lorentzian lineshape so that changes in the shift are not linearly dependent on frequency changes and so will lead to a systematic change in the shift when averaged in addition to further broadening. In this case the laser is tuned within a few MHz of resonance. The transition linewidth is $\sim 20MHz$ so relative changes in the size of the shift would be of order $\lesssim(3/20)^2/2\sim 1\%$ slightly smaller than the statistical sensitivity here. This uncertainty is reduced for shifts larger than the natural linewidth of the transition and could be eliminated by measuring the shift as a function of frequency and fitting to the expected lineshape to determine the detuning.

\section{Polarization Dependance}

For general polarizations the structure of the shifts is less straightforward and nontrivial sets of states are coupled with varying strengths. For certain limiting cases analytic results are possible. If the transition linewidth $\Gamma$ is much larger than the zeeman splitting $\Delta\omega$, and the transition rates, $\Omega$ are much smaller than $\Delta\omega$, the shifts are given by \cite{Schacht11},

$$\delta\omega_0=\Omega\left|\hat\epsilon\cdot\hat B\right|$$
$$\delta\omega_{\pm}=\frac{\Omega}{\sqrt 2}\sqrt{1-\left|\hat\epsilon\cdot\hat B\right|^2\mp \hat\sigma\cdot\hat B}$$
$$\Omega=-\frac{e E}{2 h}\frac{\left<P\left|\left|r\right|\right|S\right>}{\sqrt 3}$$

These limits are not well satisfied for these experimental conditions as $\Gamma=20 \text{MHz}$ is only slightly larger than $\Delta\omega\sim 10\text{MHz}$, and $\Omega\sim 4\text{MHz}$ is only slightly smaller than $\Delta\omega$. As a result deviation from the above behavior appear even in the next simplest case of linear polarization with $\vec E\perp\vec B$. This couples only $\Delta m=\pm 1$ transitions, leaving the $\Delta m=0$ transition uncoupled and the expected value of the shift is $\delta\omega_\pm=\Omega/\sqrt 2$, $\sqrt 2$ smaller than the $\Delta m=0$ shift for $\vec E\parallel\vec B$.

The upper plot in figure \ref{fig:MWTransitions} and table \ref{table:Shifts} show the resulting experimental shifts. The polarization angle of the 369nm light is varied with a half wave plate that happens to yield $\vec E\parallel\vec B$ when the rotation stage it is mounted in is set to $35^\circ$. $80^\circ$ then gives $\vec E\perp\vec B$. To further elucidate the structure of the polarization dependance of the shifts and intermediate angle yielding $\angle\vec E\vec B=45^\circ$ was also used and is shown in the center plot in figure \ref{fig:MWTransitions}. The intermediate shifts and evolving linewidths are apparent.

For $\vec E\perp\vec B$ the $\Delta m=\pm 1$ shifts are negative, but not as large as the $6.7/\sqrt 2=4.7 \text{MHz}$ expected, and not the same for both transitions. The shifts can be different if the 369nm light has some degree of circular polarization, and its direction of propagation has a component in the direction of $\vec B$. However, this would result in considerable $\Delta m=0$ coupling, which would result in a shifts of that transition and a corresponding broadening. To the contrary the resonance corresponding to this transition shows no shift, and is sharp, perhaps the best evidence of $\vec E\perp\vec B$.

\begin{figure}[h]
\resizebox{\columnwidth}{!}{\includegraphics{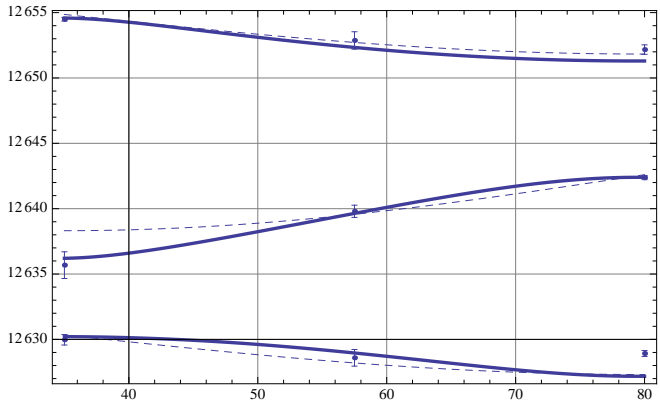}}
\caption{\label{fig:MWShifts} The frequency shift (in MHz) between the $\pi$ and $\sigma^{\pm}$ transitions of the Zeeman levels of the $^2 S_{1/2}$, F=1 as a function of the 369 nm light polarization.}
\end{figure}

The deviations from the ideal structure are more likely due to the size of the Zeeman splitting being comparable to the transition linewidth. As a result all transitions can not be driven simultaneously exactly on resonance. For example, if the 369m laser is tuned closer to the $\Delta m=+1$ transition resonance than the $\Delta m=-1$ resonance, the $S(F=1)_{m=+1}$ state will be more strongly coupled to the $P(F=0)$ state than the $S(F=1)_{m=-1}$ state and will have a larger shift. The resulting shifts in such situations can be readily calculated numerically \cite{Schacht11}, but only for the case $\Omega>>\Gamma$ which again is not satisfied here. The resulting best fit is also shown in figure \ref{fig:MWShifts}. The relative sizes of the $\delta\omega_0(\vec E\parallel\vec B)$ and the $\delta\omega_\pm(\vec E\perp\vec B)$ shifts are more consistent but the asymmetry of the $\delta\omega_\pm(\vec E\perp\vec B)$ shifts is still not reproduced. This model does not include the finite lifetime of the excited state so the transition linewidth it given only by the transition rate. Solving the full density matrix system would uncouple these two parameters and should allow for eliminating the remaining discrepancies with the observed polarization dependance of the light shift.

\begin{table}
\begin{tabular}{c|ccc}
& $\delta\omega_{-}$ & $\delta\omega_{0}$ & $\delta\omega_{+}$ \\\hline
35	& & $6.70\pm 0.12$ &\\
57.5	& $1.39\pm0.36$ & $2.58\pm0.15$ & $1.62\pm0.32$\\
80	& $1.05\pm0.22$ & & $2.32\pm0.09$
\end{tabular}
\caption{\label{table:Shifts}The Table.}
\end{table}

\section{Results and Discussion}

This preliminary experimental investigation shows that MW-Optical double resonance provides a simple means of accurately determining size of both the magnetic and optical electrical fields. At present this also provides qualitative confirmation of the relative direction of the optical polarization and the magnetic field. In this system precise determination of the polarization will require a more general numerical analysis of the system than includes the finite lifetime of the P state.

More compelling confirmation of the structure of the dependance of the light shifts on polarization will also require better frequency, intensity and polarization stability of the laser to provide sharper transition lines and reduction of possible systematic shifts from the frequency dependance of the shifts. Efforts to improve each of these is underway. Intensity stability is being improved by replacing the doubled TiSapph 370nm system with a 370nm diode laser, polarization stability by using an improved polarization maintaining fiber to deliver the 370nm light from the laser to the ion, and frequency stability by locking the laser to the ion using the cooling rate signal. A method has also been developed to measure the Zeeman splitting without the possibility of light shift perturbations.


\begin{thebibliography}{99}
\bibitem{Warrington2002} R. Warrington \emph{et al.},  Conference Digest, 156 (2002) DOI:10.1109/CPEM.2002.1034765.\\
\bibitem{Fortson1993} N. Fortson, Phys. Rev. Lett. \textbf{70}, 2383 (1993).\\
\bibitem{Mandel2010} P. Mandal \emph{et al.}, Hyperfine Int. \textbf{196}, 261 (2010).\\
\bibitem{Schauer2010} M. Schauer \emph{et al.}, Phys. Rev. A \textbf{82}, 062518 (2010).\\
\bibitem{Schacht11}M. Schacht, to be published 2012.\\
\end{thebibliography}
\end{document}